\documentclass[pra,twocolumn,superscriptaddress,nofootinbib]{revtex4}

\usepackage[usenames,dvipsnames]{color}
\usepackage{amssymb,amsmath}
\usepackage{graphicx}
\usepackage{float}
\usepackage{hyperref}
\usepackage{xfrac}

\newcommand{\ket}[1]{|{#1}\rangle}

\newcommand{\eqrf}[1]{Eq.~\eqref{#1}}

\newcommand{\op}[1]{{\hat{#1}}}
\newcommand{\GOne}{{\ensuremath{G^{(1)}}}}
\newcommand{\GTwo}{{\ensuremath{G^{(2)}}}}
\newcommand{\GammaOne}[1]{{\ensuremath{\Gamma_{#1}^{(1)}}}}
\newcommand{\GammaTwo}[1]{{\ensuremath{\Gamma_{#1}^{(2)}}}}
\newcommand{\avg}[1]{{\ensuremath{\langle #1 \rangle}}}
\newcommand{\bavg}[1]{{\ensuremath{\left\langle #1 \right\rangle}}}
\newcommand{\inp}{{\mathrm{in}}}
\newcommand{\out}{{\mathrm{out}}}
\newcommand{\amp}{{\mathrm{amp}}}
\newcommand{\N}{{\bar{N}}}
\newcommand{\fil}{{\mathrm{fil}}}
\newcommand{\eff}{{\mathrm{eff}}}
\newcommand{\GOnefil}{{\ensuremath{G^{(1)}_\fil}}}
\newcommand{\GTwofil}{{\ensuremath{G^{(2)}_\fil}}}
\newcommand{\tp}{{t_\mathrm{p}}}
\newcommand{\e}{{\mathrm e}}

\newcommand{\id}{\openone}

\begin{document}	

\title{Schemes for the observation of photon correlation functions in
  circuit QED with linear detectors}

\date{\today}

\author{Marcus P. da Silva} \affiliation{D\'epartement de Physique,
  Universit\'e de Sherbrooke, Sherbrooke, Qu\'ebec, Canada, J1K 2R1}
\author{Deniz Bozyigit} 
\affiliation{Department of Physics, ETH Z\"urich, CH-8093, Z\"urich,
  Switzerland.}
\author{Andreas Wallraff} 
\affiliation{Department of Physics, ETH Z\"urich, CH-8093, Z\"urich,
  Switzerland.}
\author{Alexandre Blais}
\affiliation{D\'epartement de Physique, Universit\'e de Sherbrooke,
  Sherbrooke, Qu\'ebec, Canada, J1K 2R1}

\begin{abstract} 
  Correlations are important tools in the characterization of quantum
  fields, as they can be used to describe statistical properties of the
  fields, such as bunching and anti-bunching, as well as to perform
  field state tomography. Here we analyse experiments
  by Bozyigit {\em et al.}~\cite{bozyigit:2010a} where
  correlation functions can be observed using the measurement records
  of linear detectors ({\em i.e.}  quadrature measurements), instead
  of relying on intensity or number detectors.  We also describe how
  large amplitude noise introduced by these detectors can be
  quantified and subtracted from the data.  This enables, in
  particular, the observation of first- and second-order coherence
  functions of microwave photon fields generated using circuit
  quantum-electrodynamics and propagating in
  superconducting transmission lines under the condition that noise is
  sufficiently low. 
\end{abstract}

\maketitle

\section{Introduction}

Field correlations are widely used in the characterization of
classical and quantum fields~\cite{mandel:2008a,walls:2008a}.  A
particular set of correlations used for such purposes are the
coherence functions of a field, as described by
Glauber~\cite{glauber:1963a,glauber:1963b,mandel:1965a}.  These
functions can be used to quantify the ability of a field to interfere
with itself, as well as to demonstrate features of quantum fields
which cannot be reproduced in a classical system. One of the most famous of
these quantum phenomena is known as {\em
  anti-bunching}~\cite{paul:1982a,davidovich:1996a}, and it is
frequently used to characterize single-photon sources in the optical
regime~\cite{kurtsiefer:2000a,yuan:2002a,keller:2004a,
  birnbaum:2005a,hijlkema:2007a}. Over the recent years
Josephson-junction based superconducting circuits, resonators and
transmission lines have emerged as a platform for performing quantum
optics experiments in the microwave
regime~\cite{wallraff:2004a,schuster:2007a,wallraff:2007a,houck:2007a,astafiev:2007a,hofheinz:2008a,
  fink:2008a,regal:2008a,beltran:2008a,hofheinz:2009a,astafiev:2010a}. While
in the optical regime coherence functions are usually measured using
an interferometer where photon number detectors are used, in the
microwave regime, linear detectors ({\em i.e.} field quadrature
measurements) are ubiquitous due to the difficulty of building
reliable photon number detectors. This raises the question of how to
measure field correlations using linear detectors. This paper answers
this question and describes the theory behind the recent experiments
performed by Bozyigit {\em et al.}~\cite{bozyigit:2010a}, where
correlations of a propagating microwave field are measured using only
linear detectors, instead of intensity detectors. While the discussion
here focuses on the measurement of first- and second-order coherence
functions of microwave fields, the analysis can be applied to any
correlation of field operators.  We note that the measurement of
correlation functions of propagating microwave fields using non-linear
(i.e.~square-law) detectors was theoretically studied in
Ref.~\cite{mariantoni:2005a}, under the assumption of negligible
correlation in the noise added by the detection chain. In practice,
these correlations turn out to be important and are discussed here.
Recent work by Menzel {\em et al.}~\cite{menzel:2010a} and Mariantoni
{\em et al.} \cite{mariantoni:2010a} is in a similar direction to the
work presented here.

The paper is organized as follows. Section~\ref{sec:coh-func} gives a
brief review of coherence functions and how they are measured with
non-linear detectors. Section~\ref{sec:lin-det} describes how field
correlations, and in particular coherence functions, can be measured
using linear detectors. Section~\ref{sec:noise} describes the effects
of noise in the experiments, and finally
Section~\ref{sec:two-sided} describes how the experimental setup can
be simplified in circuit QED experiments.

\section{Coherence functions\label{sec:coh-func}}

\begin{figure}
  \includegraphics{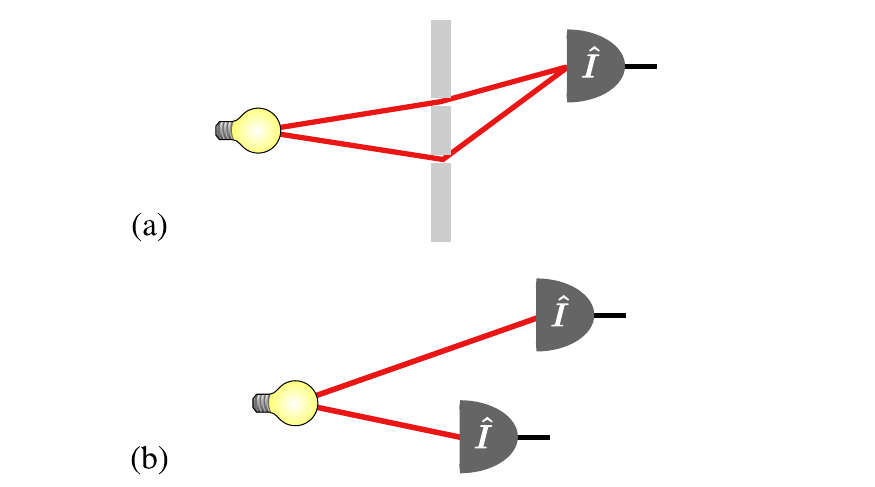}
  \caption{(Color online) Experimental setups illustrating different
    degrees of optical coherence with intensity detectors. The red
    lines represent quantum fields, and the black lines represent
    measurement records.\label{fig:gedanken}}
\end{figure}

The meaning of {\em coherence} of a field in a single frequency mode,
with corresponding annihilation operator $\op{a}$, can be understood
by considering interference experiments which use the field leaking
out of this mode.  Using a double slit, the field can be made to
travel two pathways of different lengths which terminate at a single
point-like photon detector, as depicted in
Fig.~\ref{fig:gedanken}(a). The combined field that impinges on the
detector is made up of fields originally emitted at times $t$ and
$t+\tau$ (which depend on the lengths of the paths), so that the
observed field intensity at the detector is the sum of the intensities
of the two fields plus an interference term which depends on
$\avg{\op{a}^\dag(t)\op{a}(t+\tau)}$~\cite{mandel:2008a}. Interference
effects can only be observed if this correlation is non-zero. It is
therefore natural to define
\begin{equation} 
\GOne(t,t+\tau) = \avg{\op{a}^\dag(t)\op{a}(t+\tau)},
\end{equation}
which is called the {\em first-order coherence
  function}~\cite{glauber:1963a}, as a measure of the emitted field's
potential to interfere with itself -- in other words, a measure of the
coherence of the field. One may also consider
\begin{equation} 
\GOne(\tau) = \int_{\mathcal I} dt~\GOne(t,t+\tau),
\end{equation} 
for some time interval ${\mathcal I}$ in order to obtain an expression 
that depends only on the time difference between the two paths.
If $\GOne(\tau)=0$, no interference effects can be observed for a path
difference of $c\tau$, where $c$ is the speed of light.

The measure of coherence that is most often used to distinguish
classical fields from quantum fields is the {\em second-order coherence
function} given by
\begin{align}
\GTwo(t,t+\tau) 
& = \avg{\op{a}^\dag(t)\op{a}^\dag(t+\tau)\op{a}(t+\tau)\op{a}(t)},
\end{align}
or by the integrated version
\begin{align}
\GTwo(\tau) 
& =\int_{\mathcal I} dt~\GTwo(t,t+\tau).
\end{align}
The canonical experiment which gives the physical interpretation of
$\GTwo$ is one with a single light source and two point-like
detectors, such that the field takes a time $t$ and $t+\tau$  respectively to reach
each detector, as depicted in Fig.~\ref{fig:gedanken}(b). In that case
the correlation between the detected intensities is given by
$\GTwo(t,t+\tau)$.

For classical fields, where the field operator in the expressions
above are replaced by c-numbers, one finds that
$|\GTwo(0)|\ge|\GTwo(\tau)|$, while there are quantum states of the
field that yield $|\GTwo(0)|<|\GTwo(\tau)|$ for $\tau\not=0$, a
phenomenon known as {\em anti-bunching}~\cite{paul:1982a,davidovich:1996a}. The
canonical examples of anti-bunched field states are single photon
states and squeezed states.  In the case of pulsed experiments --
where the light field state is prepared with a repetition period of $\tp$ -- one
writes instead that classical fields obey
$|\GTwo(0)|\ge|\GTwo(k\tp)|$, and that some quantum states of the
field yield $|\GTwo(0)|<|\GTwo(k\tp)|$ for $k\not=0$.  Only pulsed
experiments will be considered in the remainder of this paper, the
generalization to continuous experiments being straightforward. 

\subsection{Standard experimental setups}

\begin{figure}
  \includegraphics{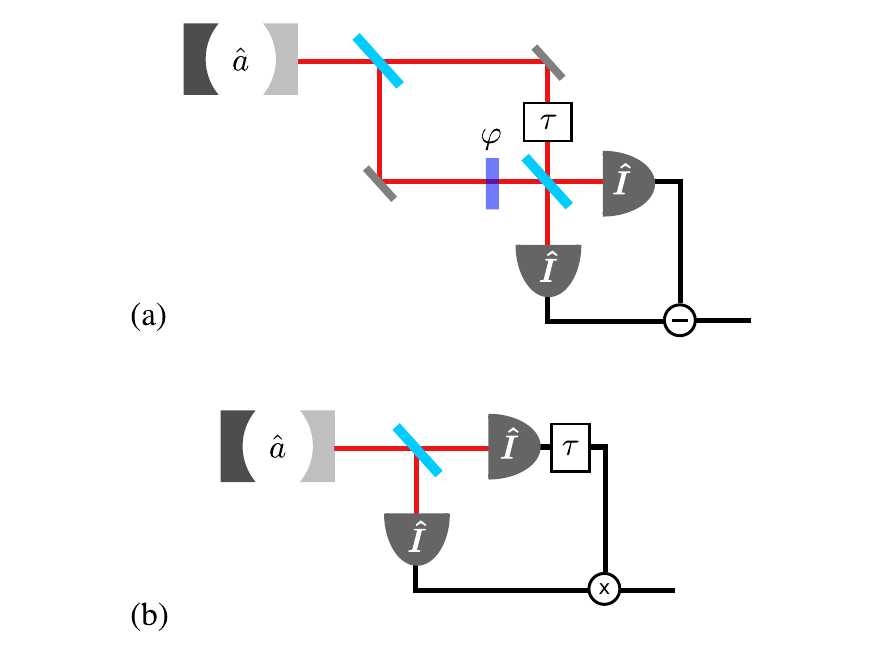}
  \caption{(Color online) Standard experiments for the observation of
    $\GOne$ and $\GTwo$ using intensity detectors: (a) a Mach-Zender
    interferometer with a variable delay $\tau$ and a variable phase
    shift $\varphi$, and (b) a Hanbury Brown and Twiss (HBT)
    interferometer with a variable delay $\tau$. The light-blue
    components are balanced beam-splitters. The cavity is taken to 
    be one-sided, with one mirror being perfectly reflective. 
   \label{fig:std-exp}}
\end{figure}

As illustrated in Fig.~\ref{fig:std-exp}, we consider experiments
where the source is a single mode of a cavity coupled to a
transmission line via a leaky mirror, a situation typical of cavity
QED~\cite{raimond:2001a,mabuchi:2002a,blais:2004a}.  In circuit QED
for example, arbitrary superpositions of a single photon and vacuum
can be prepared in the dispersive regime via Purcell
decay~\cite{houck:2007a} or by strong coupling to a qubit brought into
resonance with the cavity~\cite{bozyigit:2010a}, although details of
the state preparation are not important for the remainder of the
discussion.  The harmonic field in the cavity is associated with an
annihilation operators $\op{a}$ with the usual same-time commutation
relation $[\op{a},\op{a}^\dag]=\id$. Using input-output
theory~\cite{collett:1984a,gardiner:1985a,walls:2008a,gardiner:2004a},
one can show that $\op{a}$ is related to the modes of the transmission
line via
\begin{equation}\label{eq:one-sided-bc}
\op{b}_\out(t) = \sqrt{\kappa_b}\op{a}(t) - \op{b}_\inp(t),
\end{equation}
where $\kappa_b$ is the rate at which photons leak out of $\op{a}$,
and the input and output fields are given by
\begin{align}
\op{b}_\inp(t) & = 
{1\over\sqrt{2\pi}}\int_{-\infty}^{+\infty} d\omega\,\e^{-i\omega(t-t_0)}b(t_0,\omega)\\
\op{b}_\out(t) & = 
{1\over\sqrt{2\pi}}\int_{-\infty}^{+\infty} d\omega\,\e^{-i\omega(t-t_1)}b(t_1,\omega)
\end{align}
for transmission line modes $b(t,\omega)$ at times $t_0<t<t_1$, and
correspond to fields propagating towards or away from the cavity. 
The commutation relations of the input and output
fields are given by
\begin{align}
[\op{b}_\inp(t),\op{b}_\inp^\dag(t+\tau)]=
[\op{b}_\out(t),\op{b}_\out^\dag(t+\tau)]=\delta(\tau).
\end{align}
These definitions lead to an equation of motion for $\op{a}$ in the
interaction frame to be given, for a one-sided cavity, by
\begin{equation}\label{eq:aINOUT}
\dot{\op{a}}(t) = -{\kappa_b\over2}\op{a}(t)+\sqrt{\kappa_b}\op{b}_\inp(t).
\end{equation}
From Eq.~\eqref{eq:one-sided-bc} it is clear that the correlations of
$\op{b}_\out$ are proportional to the correlations of $\op{a}$ when
$\op{b}_\in$ is prepared in the vacuum state.  The
remainder of the discussion will focus on the observation of the
coherence functions of the output field $\op{b}_\out$ only, as they
can be taken to be equivalent to the correlation functions of
$\op{a}$. The ``out'' subscript will also be dropped when it is clear
form the context.

The state of the cavity field is taken to be prepared at times $t=k\tp$
for integer $k$ and repetition period $\tp$, and allowed to decay via
the leaky mirror as described above. The repetition period is chosen
to obey $\tp\gg 2\pi/\kappa_b$ so that the cavity can be taken to be in
equilibrium at the time of the next preparation of the cavity
field.

When working with photons in the optical frequencies, $\GOne$ 
is usually observed using a Mach-Zender
interferometer with a variable delay of $\tau$ in one of the
branches~\cite{bachor:2004a}, as depicted in
Fig.~\ref{fig:std-exp}(a). The difference between the intensities in
the photo-current detectors can yield the real or the imaginary part of
$\GOne$, depending on the phase shift $\varphi$ in the lower
branch. The standard approach to the observation of the $\GTwo$ is to use a
Hanbury Brown and Twiss (HBT) interferometer~\cite{bachor:2004a},
which is illustrated in Fig.~\ref{fig:std-exp}(b). In order to observe
$\GTwo$, one simply measures the correlations between the
photo-currents of the two detectors.

Both these setups rely on field intensity detectors, which give
information about the number of photons, and thus can be modeled by
non-linear quantum optical interactions\footnote{Linearity in this
  sense refers to the representability of the Heisenberg picture
  evolution by a linear transformation of creation and annihilation
  operators for all times.}.  Low-noise intensity detectors for
optical fields are common, and although non-linear detectors have been
demonstrated in the microwave regime~\cite{gabelli:2004a,houck:2007a}
(albeit with higher noise levels than in optics), the main motivation
for this paper is to illustrate how the coherence functions of
microwave fields in circuit QED may be measured through the use of
linear detectors only.

\section{Linear detectors\label{sec:lin-det}}

Field quadrature measurements of microwave signals is a standard
technique~\cite{pozar:2004a} which has been applied very successfully
to quantum electrical circuits in the recent years to demonstrate, for
example, new regimes of cavity QED~\cite{wallraff:2004a},
high-contrast detection of qubit states~\cite{lupascu:2006a}, photon
states~\cite{schuster:2007a}, and nanomechanical oscillator
states~\cite{teufel:2009a}. Since field quadrature operators are
fundamentally different from number operators, different experimental
setups are required in order to measure the coherence functions
$\GOne$ and $\GTwo$. Grosse {\em et al.}~\cite{grosse:2007a} have
demonstrated how a HBT interferometers can be modified to measure
$\GTwo$ using field quadratures instead of intensity
measurements. Here we analyse similar
experiments~\cite{bozyigit:2010a}, and consider generalizations and
simplifications which exploit features of circuit QED, while at the
same time considering the large added noise due to the HEMT amplifiers
currently required for measurement in this system.

The details of the implementation of quadrature operator measurements
in the microwave regime are different from the standard optical
implementation. In particular, homodyne detection in the microwave
regime is performed via mixing instead of beam splitting~\cite{pozar:2004a}. 
For simplicity, we will however consider the optical analogues
of the devices we discuss. Common non-idealities in the microwave
regime, such as weak thermal states instead of vacuum inputs, can be
treated straightforwardly by considering different input states, and
thus do not change the analysis significantly.

\begin{figure}
\includegraphics{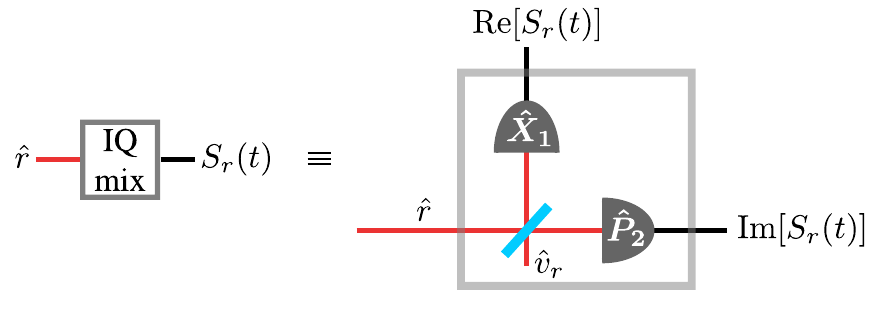}
\caption{
(Color online) Optical analog of an IQ-mixer as an 8-port homodyne detector. The field $\op{r}$ is fed
into a balanced beam splitter along with vacuum $\op{v}_r$. The
$\op{X}$ quadrature of one of the beam splitter outputs is measured
while the $\op{P}$ quadrature of the other output is measured. The classical 
outcomes are the real and imaginary parts of the complex envelope $S_r(t)$.
\label{fig:iq-mixer}}
\end{figure}

The measurement of both quadratures of a propagating field, realized
in optics through 8-port homodyne~\cite{schleich:2001a} or heterodyne
detection, is performed by an {\em IQ mixer} in the microwave
regime~\cite{pozar:2004a}.  The symbol for the IQ mixer, and its
description in terms of its optical analogue are depicted in
Fig.~\ref{fig:iq-mixer}. The input is any propagating quantum field
with annihilation operator $\op{r}$, which may stand for any
propagating field considered in this paper.  The outputs are
quadrature measurements of the superpositions of the $\op{r}$ field
with a mode $\op{v}_r$ in the vacuum state, where
$[\op{r},\op{v}_r^\dag]=0$. These outputs are labeled $X_1$ and $P_2$
to emphasize that the measurements are made on different commuting
modes, and correspond to the in-phase component and the quadrature
component of the measurement respectively.

Finally, it is important to note that, for most circuit QED
experiments, only averages of these quadratures over many realizations
of the experiment are measured. Here, however, we are interested in
experiments where the full time records of these quadratures are
recorded, for each realizations of the
experiments~\cite{bozyigit:2010a}. Based on these full records, any
averages or correlation functions can be reconstructed, as is
discussed in the next sections.

\subsection{Complex envelope}

Given the two classical outputs $X_1(t)$ and $P_2(t)$, it is useful to define the
complex envelope $S_r(t)$ of $\op{r}$ as
\begin{equation}
S_r(t) = X_1(t) + i P_2(t),
\end{equation}
which is a random c-number due to the dependence on the measurement
records of the quadratures.  Noting that
\begin{align}
\avg{X_1(t)} 
&= \bavg{{\op{r}_1+\op{r}_1^\dag\over\sqrt{2}}} 
= \avg{\op{X}_r(t)}+\avg{\op{X}_v(t)} \\
\avg{P_2(t)}
&= -i\bavg{{\op{r}_2-\op{r}_2^\dag\over\sqrt{2}}} 
= \avg{\op{P}_r(t)}-\avg{\op{P}_v(t)},
\end{align}
one may write that $\avg{S_r(t)}=\avg{\op{S}_r(t)}$ where the complex
envelope operator $\op{S}_r$ is defined by
\begin{equation}
\op{S}_r(t) \equiv \op{r}(t)+\op{v}^\dag_r(t) = \op{X}_1(t)+i\op{P}_2(t).\label{eq:comp-env-op}
\end{equation}
In order to simplify the remainder of the calculations, it is convenient to
define $\op{S}_r$ in this manner instead of using the quadrature operators
explicitly. 

Given that the mode $\op{v}_r$ is in the vacuum state, the expression
for the expectation values take simple forms. The presence of the
vacuum mode $\op{v}_r$ is indeed important as it leads to the
commutation relation
\begin{equation}
[\op{S}_r(t),\op{S}_r^\dag(t')]=0,
\end{equation}
implying that $\op{S}_r$ is normal and therefore diagonalizable. Since
$\op{S}_r$ is described by the sum of the commuting operators
$\op{X}_1$ and $\op{P}_2$, its eigenvalues are given by the sum of the
eigenvalues of these operators for any fixed eigenvector. This
corresponds to the measurement record $S_r$ of $\op{S}_r$ being simply
the sum of the measurement records $X_1$ and $P_2$, as claimed
earlier. Note that this does {\em not} imply that both quadratures of
$\op{r}$ can be measured simultaneously without back-action.

Since $\op{S}_r$ and $\op{S}^\dag_r$ commute at all times, arbitrary
correlations of these operators, like the correlations of their
measurement records, do not depend on operator ordering. Therefore,
\begin{equation}
\avg{(S_r^*)^m S_r^n} 
= \avg{(\op{S}_r^\dag)^m \op{S}_r^n} 
= \avg{(\op{r}^\dag+\op{v}_r)^m (\op{r}+\op{v}_r^\dag)^n},
\end{equation}
independently of the ordering of the terms. However, in order to
reduce these expressions to a correlation function of $\op{r}$ alone,
one must rewrite the expression such that the $\op{v}_r$ modes are in
normal ordering in order to immediately evaluate the expectation
values, leading to 
\begin{equation}
\avg{(S_r^*)^m S_r^n} 
= \avg{(\op{S}_r^\dag)^m \op{S}_r^n} 
= \avg{\op{r}^n (\op{r}^\dag)^m},
\end{equation}
or in other words, correlations of the complex envelope of a field
correspond to anti-normally ordered correlations of the field
operator, under the assumption that the $\op{v}_r$ mode is prepared in
the vacuum. In this case the measurement of $\op{S}_r$ is described by
the Husimi-Kano $Q$ function which is known to give access to
anti-normally ordered same-time
correlations~\cite{husimi:1940a,kano:1965a,schleich:2001a}. Similar
results hold for multi-time correlations.

It is important to note that, while with this approach arbitrary
correlations can be evaluated, the number of statistical samples
needed to obtain a desired precision in the estimate grows as the
noise power raised to the desired correlation order (see
Appendix~\ref{app:stat-error} for details).  In practice, this limits
the order of the correlations measured with current amplifier noise levels
due to the large number of repetitions of the experiment needed to
obtain reasonable error bars. Use of quantum limited amplifiers would
greatly improve the situation~\cite{regal:2008a,beltran:2008a,clerk:2010a}.

\subsection{$\GOne$ observation}

\begin{figure}
\includegraphics{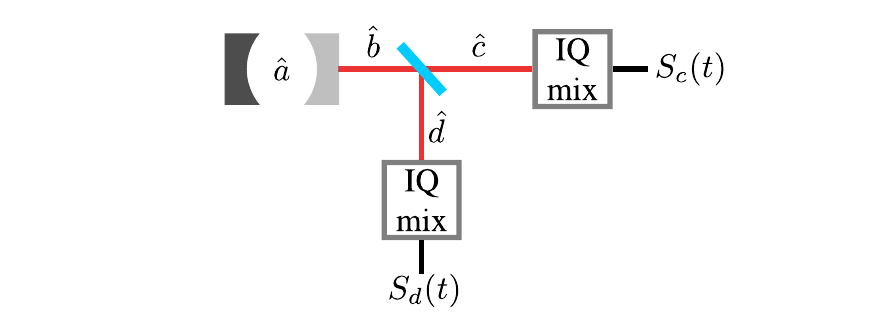}
\caption{(Color online) Hanbury Brown and Twiss interferometer with complex envelope
  measurement used to measure the first-order coherence function
  $\GOne$. The output of the cavity field is separated by an on-chip
  beam splitter and the resulting fields measured by an IQ-mixer.
\label{fig:hbt-g1}}
\end{figure}

As depicted in  Fig.~\ref{fig:hbt-g1}, with IQ-mixers, the first order correlation
function $\GOne$ can be measured from the outputs of a HBT interferometer.
Because of the unitary of the beam splitter and the presence of a vacuum port,
the complex envelope operators of the outputs labeled $\op{c}$ and $\op{d}$
commute.

The auto-correlation of one of the complex envelopes, say $S_c(t)$,
is given by
\begin{equation}
\GammaOne{\alpha}(t,t+\tau) 
= \avg{\op{S}^\dag_c(t)\op{S}_c(t+\tau)} 
= \delta(\tau)+{1\over2}\avg{\op{b}^\dag(t)\op{b}(t+\tau)},
\end{equation}
while the cross-correlation between the complex envelopes is 
\begin{equation}
\GammaOne{\beta}(t,t+\tau) 
= \avg{\op{S}^\dag_c(t)\op{S}_d(t+\tau)} 
= {1\over2}\avg{\op{b}^\dag(t)\op{b}(t+\tau)},
\end{equation}
where we have used the fact that the expectation values of all the vacuum modes
are zero. Thus the first-order coherence function $\GOne$ of the $\op{b}$ field
is immediately accessible from cross-correlations of the
complex envelopes in a modified HBT interferometer via
\begin{equation}
\begin{split}
\GOne(t,t+\tau) 
& = 2\GammaOne{\alpha}(t,t+\tau)-2\delta(\tau),\\
& = 2\GammaOne{\beta}(t,t+\tau),
\end{split}
\end{equation}
up to non-idealities, such as amplifier noise, which will be treated later.
The exact expressions for $\GOne{}$ of the states prepared in
Ref.~\cite{bozyigit:2010a} are given in Appendix~\ref{app:g1-g2}.

Although the divergence of the $\delta$ functions may appear
problematic, in reality due to the finite bandwidth of the experiments
these delta functions are replaced by smooth bounded
functions, while the coherence functions are distorted by a
convolution kernel which preserves the relative heights of the peaks
in the experiment. This results in the filtered correlation functions
\begin{equation}
\begin{split}
\GammaOne{\alpha,\fil}(\tau) & = 
{1\over 2}\GOnefil(\tau)+f_\eff(\tau),\\
\GammaOne{\beta,\fil}(\tau) & = 
{1\over 2}\GOnefil(\tau).
\end{split} 
\end{equation}
where $\GOnefil(\tau)=\GOne(\tau)\ast f_\eff(\tau)$ and $f_\eff$ is a
function describing the effective action of the filter
(see Appendix~\ref{app:filtering} for details). 

\subsection{$\GTwo$ observation}

The expressions needed to measure the second-order coherence
function from the complex envelopes can be constructed by inspection 
from Eq.~\eqref{eq:comp-env-op}. Depending on which factors are
taken to be complex conjugates or to be displaced in time by $\tau$,
different correlations can be used to extract information about
$\GTwo$. One such choice is
\begin{multline}
\GammaTwo{\alpha}(t,t+\tau) 
= \avg{\op{S}_c^\dag(t) \op{S}_d^\dag(t+\tau) \op{S}_d(t+\tau)
  \op{S}_c(t)} \\
= 
\delta^2(0)
+{1\over2}\avg{\op{b}^\dag(t)\op{b}(t)}\delta(0)
+{1\over2}\avg{\op{b}^\dag(t+\tau)\op{b}(t+\tau)}\delta(0) \\
+{1\over4}\avg{\op{b}^\dag(t) \op{b}^\dag(t+\tau) \op{b}(t+\tau)\op{b}(t)},
\end{multline}
so that $\GTwo$ can be obtained immediately via
\begin{multline}
\GTwo(t,t+\tau) = 4\GammaTwo{\alpha}(t,t+\tau) - 2\GOne(t,t)\delta(0) \\
- 2\GOne(t+\tau,t+\tau)\delta(0) - 4\delta^2(0).
\end{multline}
Another choice that leads more directly to $\GTwo$ is
\begin{equation}
\begin{split}
\GammaTwo{\beta}(t,t+\tau) 
&= \avg{\op{S}_c^\dag(t) \op{S}_c^\dag(t+\tau) \op{S}_d(t+\tau)
  \op{S}_d(t)}, \\ 
&= {1\over4}\avg{\op{b}^\dag(t)\op{b}^\dag(t+\tau)\op{b}(t+\tau)\op{b}(t)},
\end{split}
\end{equation}
so that
\begin{equation}
\GTwo(t,t+\tau)
= 4 \GammaTwo{\beta}(t,t+\tau).
\end{equation}
As described earlier, the divergence of the $\delta$ functions is
taken care of by filtering in a realistic experiment.  The main
distinction between these two approaches of measuring the second-order
coherence functions is how they are affected by noise in the
experiment, as is discussed in the next section.

\section{Rejection and subtraction of noise
  \label{sec:noise}}

The amplitude of microwave signals in a superconducting quantum
circuit is small enough that amplifiers are essential for their
observation, and so in a realistic experiment, the field is amplified
\emph{before} mixing. Using the Haus-Caves description of a quantum
amplifier~\cite{caves:1982a,gardiner:2004a,clerk:2010a}, an input
operator $\op{c}$ and an output operator $\op{c}_\amp$ for a
phase-preserving amplifier with gain $g_c$ are related by
\begin{equation}\label{eq:camp}
\op{c}_\amp = \sqrt{g_c}~\op{c} + \sqrt{g_c-1}\op{h}_c^\dagger,
\end{equation}
where $\op{h}_c$ is an added noise mode.

It is clear that if $g_c>1$ there will be added noise due to
amplification, even at zero temperature. However, for thermal white
Gaussian noise, one finds that all odd order moments vanish. As a result,
the first moments of quadrature fields are not affected by this
amplifier noise, just as they are not affected by vacuum noise. The
contributions from other moments may be non-zero, however, and must be
accounted for. For simplicity, we only consider the case of Gaussian
white noise here, but similar results follow straightforwardly for
general noise as long as the noise is independent of the inputs. Since
the noise moments can be extracted from experimental data, the
assumption of Gaussian noise is not essential.

The noise modes from different amplifiers are taken to commute, but in
general they may be correlated.  While the noise is normally taken to
come from the amplification~\cite{caves:1982a,gardiner:2004a,
  clerk:2010a}, formally one may also take $\op{h}_c$ to include
thermal noise from other sources, such as the vacuum ports of the
IQ-mixer and of the beam-splitter, with only minor modifications. 
Here $\op{h}_c$ is taken to have a commutator
$[\op{h}_c(t),\op{h}_c^\dag(t+\tau)]=\delta(\tau)$ in order to
preserve the bosonic commutation relations of the amplified signals 
$\op{c}_\amp$, and auto-correlation
$\avg{\op{h}_c^\dag(t)\op{h}_c(t+\tau)}=\N_c\delta(\tau)$.  The noise
sources are assumed to be independent of the inputs, so that
$[\op{c},\op{h}_c]=[\op{c},\op{h}_c^\dag]=0$, and
$\avg{\op{c}\op{h}_c}=\avg{\op{c}\op{h}_c^\dag}=0$. The correlations
between $\op{h}_c$ and the noise mode $\op{h}_d$ from the other
amplifier in the experiments described here is taken to be
$\avg{\op{h}_c(t)\op{h}_d^\dag(t+\tau)}=\N_{cd}\delta(\tau)$ while
$\avg{\op{h}_c(t)\op{h}_d(t+\tau)}=0$.


Using this noise model, one can calculate the
different correlations $\GammaOne{\alpha,\beta}$ using the amplified
modes, resulting in
\begin{equation}
\begin{split}
\GammaOne{\alpha,\amp}(t,t+\tau) &= 
{g_c\over 2}\GOne(t,t+\tau)+(\N_{c}+g_c)\delta(\tau),\\
\GammaOne{\beta,\amp}(t,t+\tau) &= 
{\sqrt{g_cg_d}\over 2}\GOne(t,t+\tau)+\N_{cd}\delta(\tau),
\label{eq:bs-g1-alpha-beta}
\end{split} 
\end{equation}
in the unfiltered case.

Since the thermal noise in the amplifiers is independent of the
inputs, a steady-state experiment with the input mode $\op{b}$ in the
vacuum state can be used to estimate the noise strengths and subtract
the corresponding terms from $\GammaOne{\alpha,\beta,\amp}$ to obtain an estimate of
$\GOne$. When the noise cross-correlation $\N_{cd}$ is expected to
be zero or negligible compared to the noise auto-correlations $\N_c$
and $\N_d$, the approach to the estimation of $\GOne$ based on
$\GammaOne{\beta,\amp}$ provides {\em noise rejection} without 
additional post-processing.

The second-order coherence function for the amplified fields has
similar properties. One finds
\begin{widetext}
\begin{multline}
\GammaTwo{\alpha,\amp}(t,t+\tau) = 
{g_cg_d\over4}\GTwo(t,t+\tau)+{g_c\over2}\delta(0)[g_d+\N_d]\GOne(t,t)
+{g_d\over2}\delta(0)[g_c+\N_c]\GOne(t+\tau,t+\tau)\\
+{\sqrt{g_cg_d}\over2}\N_{cd}\delta(\tau)
[\GOne(t+\tau,t)+\GOne(t,t+\tau)]
+[g_c\N_d+g_cg_d+g_d\N_c]\delta^2(0)
+\avg{\op{h}_d^\dag(t+\tau)\op{h}_c^\dag(t)\op{h}_c(t)\op{h}_d(t+\tau)},
\label{eq:bs-g2-alpha}
\end{multline}
while
\begin{multline}
\GammaTwo{\beta,\amp}(t,t+\tau) = 
{g_cg_d\over4}\GTwo(t,t+\tau)
+\avg{\op{h}_d^\dag(t+\tau)\op{h}_d^\dag(t)\op{h}_c(t)\op{h}_c(t+\tau)}
\\
+{\sqrt{g_cg_d}\over2}\N_{cd}
\left[
\delta(\tau)\GOne(t+\tau,t)
+\delta(0)\GOne(t+\tau,t+\tau)
+\delta(0)\GOne(t,t)
+\delta(\tau)\GOne(t,t+\tau)
\right],
\label{eq:bs-g2-beta}
\end{multline}
\end{widetext}
where all odd moments of the noise modes where taken to be zero (if
such an assumption cannot be made, similar expressions involving the
odd moments are easily derived but are omitted here for brevity).  The
recovery of the second-order coherence function from noisy signals is
clearly more involved, but requires only the estimation of first-order
coherence functions, as well as two and four-point noise correlations
in an experiment where the input mode $\op{b}$ is prepared in the
vacuum. Since the filters are taken to be linear and time-invariant,
$\GTwofil$ is a scaled and distorted version of $\GTwo$, preserving
the relative heights of the peaks, so that the non-classical
properties of the field can still be verified.  Once again we see that
$\GammaTwo{\beta,\amp}$ provides a more direct estimation of $\GTwo$
by rejecting contributions from uncorrelated noise up to four-point
noise correlations.

\section{Two-sided cavities\label{sec:two-sided}}

\begin{figure}
\includegraphics{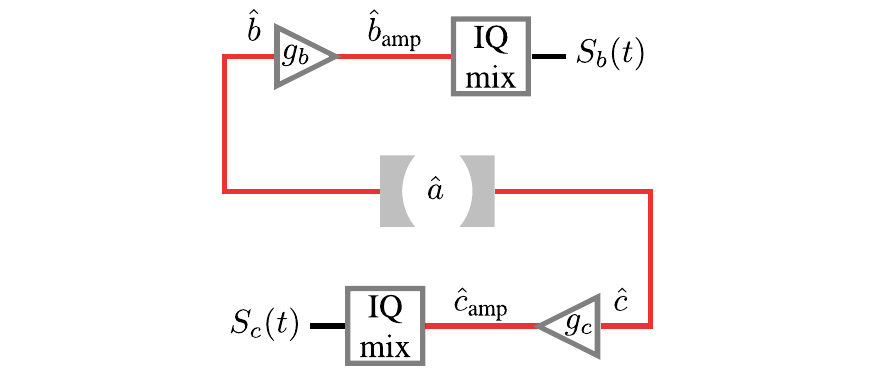}
\caption{(Color online) The setup for the observation of coherence functions
  using a two-sided cavity.
\label{fig:two-sided}}
\end{figure}

Strictly speaking, the beam splitter is not necessary for the
observation of the coherence functions described above. If one
considers a two-sided cavity, illustrated in Fig.~\ref{fig:two-sided}, 
the correlations between the cavity
outputs behave in a manner similar to the outputs of the beam splitter
in the HBT interferometers. In particular, using causality as well as
the boundary conditions of the input and output fields of the
two-sided cavity, Appendix~\ref{app:commutation} shows that
\begin{equation}
[\op{b}_\out(t),\op{c}_\out^\dagger(t')]=0,
\end{equation}
where $\op{c}_\out$ is defined in a manner analogous to $\op{b}_\out$,
with a mirror leakage rate $\kappa_c$, and an amplifier with gain
$g_c$ being applied before mixing and measurement. 
It is thus possible to measure the complex envelopes of the two cavity
outputs and calculate the correlations in the same manner as in the
modified HBT setup without the need for an additional beam
splitter. This can lead to simpler and smaller experimental setups, as
beam splitters in the microwave regime can occupy a significant area
in coplanar devices.

Calculating the correlations using the two cavity outputs
$\op{b}_\out$ and $\op{c}_\out$ one finds
\begin{widetext}
\begin{align}
\GammaOne{\alpha,\amp}(t,t+\tau) &= 
{\kappa_cg_c}\GOne(t,t+\tau)+(\N_{c}+g_c)\delta(\tau),\\
\GammaOne{\beta,\amp}(t,t+\tau) &= 
{\sqrt{\kappa_b\kappa_c}\sqrt{g_cg_d}}\GOne(t,t+\tau)+\N_{bc}\delta(\tau),\\
\begin{split}
\GammaTwo{\alpha,\amp}(t,t+\tau) & = 
{g_bg_c\kappa_b\kappa_c}\GTwo(t,t+\tau)
+[\N_b+g_b]\delta(0)g_c\kappa_c\GOne(t,t)
+[g_c+\N_c]\delta(0) {g_b\kappa_b}\GOne(t+\tau,t+\tau)\\
& +{\sqrt{g_bg_c}\sqrt{\kappa_b\kappa_c}}\N_{bc}\delta(\tau)
[\GOne(t+\tau,t)+\GOne(t,t+\tau)]\\
& +[g_b\N_c+g_bg_c+g_c\N_b]\delta^2(0)
+\avg{\op{h}_c^\dag(t+\tau)\op{h}_c^\dag(t)\op{h}_b(t)\op{h}_c(t+\tau)},
\end{split}
\\
\begin{split}
\GammaTwo{\beta,\amp}(t,t+\tau) & = 
{g_bg_c\kappa_b\kappa_c}\GTwo(t,t+\tau)
+\avg{\op{h}_b^\dag(t+\tau)\op{h}_b^\dag(t)\op{h}_c(t)\op{h}_c(t+\tau)}
\\
& +{\sqrt{g_bg_c}\sqrt{\kappa_b\kappa_c}}\N_{bc}
\left\{
\delta(\tau)[\GOne(t+\tau,t)+\GOne(t,t+\tau)]
+\delta(0)[\GOne(t+\tau,t+\tau)+\GOne(t,t)]
\right\},
\end{split}
\end{align}
\end{widetext}
where $\GOne$ and $\GTwo$ are now the coherence functions of the
cavity field $\op{a}$ instead of the cavity output fields, leading to
the introduction of additional factors which depend on the cavity
leakeage rates $\kappa_{b,c}$. These expressions are directly analogous to
Eqs.~\eqref{eq:bs-g1-alpha-beta}, \eqref{eq:bs-g2-alpha}, and
\eqref{eq:bs-g2-beta}.

\section{Summary}

We have analysed experiments for the measurement of field correlations
using only field quadrature detectors and in the situation where the
full record of many repetitions of the experiment are available.  The
combination of the quadrature measurements into complex envelopes
gives direct access to anti-normally ordered field correlations. While
re-ordering of the operators in the correlations and the use of
phase-preserving amplifiers introduces additional noise into these
measurements, we demonstrated that the noise can be accounted for and
subtracted in order to reveal only the field correlations of interest.
Although there are indications that the number of statistical samples
scales exponentially with the order of the correlation function, the
measurement of low order correlations is possible for current
amplifier noise levels.

\section{Acknowledgments}

M.P.S. was supported by a NSERC postdoctoral fellowship.  D.B. and
A.W. were supported by ERC and ETHZ.  A.B. was supported by NSERC, 
the Alfred P. Sloan Foundation and is a CIFAR Scholar.

\appendix

\section{Statistical error on correlation function estimates\label{app:stat-error}}

In order to estimate the minimal number of repetitions of the
experiment which must be performed to extract a given correlation function,
consider the product of
uncorrelated Gaussian random variables with zero mean and identical
variances $\sigma^2$. These random variables correspond to the
measurements of different outputs at steady-state after the cavity
state has decayed, and the variances are given by the noise power of
the measurement record (including vacuum noise). In order to
illustrate the argument, we consider real valued random variables
$V_i$ first, and generalize to complex valued random variables $C_i$.
Since these random variables are uncorrelated, it follows that
$\avg{V_1V_2\cdots V_m}=0$. However, given a finite number of
statistical samples, the sample average $\overline{V_1V_2\cdots V_m}$
will deviate from zero due to statistical fluctuations. Signal
features which are comparable with the typical size of these
fluctuations cannot be reliably observed. As the typical size of these
fluctuations decreases with the increasing number of repetitions, this
is in principle not a fundamental problem.

In order to estimate the number of samples needed for the reliable
estimation of two-point correlations, consider the product of two
Gaussian random variables.  The characteristic function of this
product is given by
\begin{align}
\phi(U) & = 
\int_{\mathbb R^3} dv_1\,dv_2\,du\, p_1(v_1) p_2(v_2) \delta(v_1 v_2 -
u) \e^{-i U u}\\
&={1\over\sqrt{1 + U^2 \sigma^4}}
\end{align}
The characteristic function of the average of $R$ samples is given by
\begin{equation}
\phi_R(U) = \left[\phi\left({U\over n}\right)\right]^R.
\end{equation}
Given some error $\epsilon>0$ and a number of repetitions $R$, the
probability that the sample average $\overline{V_1V_2}$ obeys
$-\sfrac{\epsilon}{2}<\overline{V_1V_2}<\sfrac{\epsilon}{2}$ is given
by the integral of the inverse Fourier transform of $\phi_R(U)$ over
this range and simplifies to
\begin{equation}
\Pr\left(|\overline{V_1V_2}|<\sfrac{\epsilon}{2}\right)=
{1\over2\pi}
\int_{-\infty}^{+\infty}dU\,\phi_R(U) 
{ \sin\sfrac{\epsilon U}{ 2} \over \sfrac{U}{2}}
\end{equation}
which can be evaluated by numerically. Thus it is
straightforward to calculate the number of repetitions $R$ required to
observe a feature larger than $\epsilon$ with confidence 
$\Pr\left(|\overline{V_1V_2}|<\sfrac{\epsilon}{2}\right)$.

Another approach that provides a looser bound, but is more readily
generalized to higher order correlations, is based on
Chebyshev's inequality~\cite{rosenthal:2009a}. The variance of the
product of independent random variables with zero mean is the product
of the variances of each of the random variables. In the case of $R$
samples of the product of $m$ independent random variables $V_i$ one
finds that
\begin{equation}
\Pr\left( 
\left|\overline{V_1V_2\cdots V_m}\right| < 
\sfrac{\epsilon}{2} \right)
> 1 - 4 \tfrac{\sigma^{2m}}{ R \epsilon^2}.
\end{equation}
Note that in order to obtain this bound no assumption
was made about the form of distribution of the random variables, other
than the fact that the random variables are independent. Solving for
$R$ one obtains the worst-case upper bound
\begin{equation}
R < {4 \sigma^{2m} \over \epsilon^2 [1-\Pr\left( 
\left|\overline{V_1V_2\cdots V_m}\right| < 
\sfrac{\epsilon}{2} \right)]},
\end{equation}
which makes clear the exponential relationship between the order of
the correlation and the number of samples needed to have a
statistical error of less than $\sfrac{\epsilon}{2}$ with 
some fixed probability.

In order to generalizing this to complex-valued random variables $C_i$ --
where the real and imaginary parts of $C_i$ are independent with
variance $\sigma$, and the $C_i$ are mutually independent --
simply consider the real and imaginary parts of the correlations
separately. In that case, because a larger number of terms contribute
to the real and imaginary parts of the correlation, the variance has
a larger bound, and one finds
\begin{equation}
R < 
{ 8 m\sigma^{2m} \over \epsilon^2 
  \Pr(\text{error})},
\end{equation}
where $\Pr(\text{error})$ is the probability that the
absolute value of the real or imaginary parts of
$\overline{C_1C_2\cdots C_m}$ are greater than $\sfrac{\epsilon}{2}$.

There is no indication that taking into account the Gaussian
statistics of the random variables leads to better scalings. Thus the
ratio of the number of statistical samples needed to estimate
$\GTwo{}$ vs. $\GOne{}$ for some fixed noise variance and desired
accuracy is at worse proportional to the noise power in the
experiments. As a result the noise added by the amplifier can be the
crucial element in determining the feasibility of a correlation
function experiment. It becomes even more important for higher
order correlations, where the number of samples depends on the noise
power raised to some larger exponent.

\section{Coherence functions for states with 
  at most one photon\label{app:g1-g2}}

In the experiments described here~\cite{bozyigit:2010a}, the cavity is
periodically prepared in the state $\alpha\ket{0}+\beta\ket{1}$, with
a period $\tp$ such that $\kappa\tp\gg1$. This ensures that, to a very
good approximation, the cavity returns to the vacuum state before the
superposition is prepared again.

The coherence functions can be calculated straightforwardly via their
definitions in terms of the field correlations, while the correlations
can be calculated by solving the Heisenberg equations of motion for
the cavity field, and using the quantum regression
theorem~\cite{lax:1963a,carmichael:1993a,gardiner:2004a}.  This
procedure can be greatly simplified by noting that, if $t$ and
$t+\tau$ are in different preparation periods, then
\begin{equation}
  \avg{\op{a}^\dagger(t)\op{a}(t+\tau)}=
  \avg{\op{a}^\dagger(t)}\avg{\op{a}(t+\tau)},
\end{equation}
and
\begin{multline}
\avg{\op{a}^\dagger(t)\op{a}^\dagger(t+\tau)\op{a}(t+\tau)\op{a}(t)}=\\
\avg{\op{a}^\dagger(t)\op{a}(t)}\avg{\op{a}^\dagger(t+\tau)\op{a}(t+\tau)},
\end{multline}
due to the assumption $\kappa\tp\gg1$.

In the case were $t$ and $t+\tau$ are between $k\tp$ and $(k+1)\tp$
for some integer $k$, one finds that
\begin{gather}
\avg{\op{a}^\dagger(t)\op{a}(t+\tau)}  = 
\avg{\op{n}(0)} \e^{-\kappa(t-k\tp-\tau/2)},\\
\avg{\op{a}^\dagger(t) \op{a}^\dagger(t+\tau) \op{a}(t+\tau) \op{a}(t)
} = \avg{\op{a}^\dagger\op{a}^\dagger\op{a}\op{a}} \e^{-\kappa(2t-2k\tp+\tau)},
\end{gather}
while if $t$ and $t+\tau$ are in different preparation periods 
starting at $k\tp$ and $(k+l)\tp$, one finds that
\begin{gather}
\avg{\op{a}^\dagger(t)\op{a}(t+\tau)} = 
|\avg{\op{a}(0)}|^2 \e^{-\kappa[t-k\tp-(\tau-l\tp)/2]},\\
\avg{\op{a}^\dagger(t) \op{a}^\dagger(t+\tau) \op{a}(t+\tau) \op{a}(t)
} = \avg{\op{n}(0)}^2 \e^{-\kappa(2t-2k\tp+\tau-l\tp)}.
\end{gather}
After integration over $t$, the first-order coherence function can be
shown to be well approximated by
\begin{multline}
\GOne(\tau) = 
{1\over\kappa} \avg{\op{n}(0)} \e^{-\kappa|\tau|/2} \\
+ {1\over\kappa} |\avg{\op{a}(0)}|^2 \sum_{l\not=0} \e^{-\kappa|\tau-l\tp|/2}.
\end{multline}
This can be interpreted as a series of time-shifted copies of
$\e^{-\kappa|\tau|/2}$, where the peak centered at $\tau=0$ has a
height equal to $\avg{n(0)}$, while the peaks centered at non-zero
multiples of $\tp$ have a height equal to $|\avg{a(0)}|^2$.

Under similar assumption, the second order correlation function
can be shown to be well approximated by
\begin{multline}
\GTwo{}(\tau) = 
{1\over\kappa} \avg{\op{a}^\dagger(0) \op{a}^\dagger(0) \op{a}(0)
  \op{a}(0)} \e^{-\kappa|\tau|}\\
+{1\over\kappa} \avg{\op{n}(0)}^2 \sum_{l\not=0} \e^{-\kappa|\tau-l\tp|},
\end{multline}
such that the center peak has a height proportional to
$\avg{\op{a}^\dag(0) \op{a}^\dag(0) \op{a}(0) \op{a}(0)}$ while the
other peaks have heights proportional to $\avg{\op{n}(0)}^2$.

For the superpositions of vacuum and a single photon considered in
\cite{bozyigit:2010a}, we find that
\begin{align}
\avg{\op{n}(0)} & = |\beta|^2,\\
|\avg{\op{a}(0)}|^2 & = |\alpha|^2|\beta|^2,\\
\avg{\op{a}^\dag(0) \op{a}^\dag(0) \op{a}(0) \op{a}(0)} & = 0,\\
\avg{\op{n}(0)}^2 & = |\beta|^4,
\end{align}
indicating that the center peak of $\GTwo{}$ is abscent,
while the other peaks are non-zero, which is a signature of the purely
quantum effect known as anti-bunching~\cite{paul:1982a,davidovich:1996a}.

\section{Filtering\label{app:filtering}}

The finite bandwidth of the detection chain can be modeled by
considering the insertion of a bandpass filter in an ideal (infinite
bandwidth) detection chain. In order to calculate the effect of
filtering on correlation functions one can consider a general
framework which describes what happens to multi-time, multi-channel
correlations when measurement signals are filtered. Assume a system
with $n$ channels where each channel is filtered individually. One can
write the filtered outcome of each channel $S_{\fil,i}$ in terms of
the input signal $S_i$ and the filter function $f_i$ by using the
relations for linear time-invariant systems~\cite{proakis:2006a}
\begin{equation}
\begin{split}\label{eq:defFilteredSignal}
S_{\fil,i}(t_i) &= f_i(t_i) \ast S_i(t_i) \\
&= \int_{-\infty}^{+\infty} f_i(\tau_i) S_i(t_i-\tau_i) d\tau_i.
\end{split}
\end{equation}
Each channel has a separate time variable $t_i$ to capture the case of
multi-time correlations. This also clarifies with respect to which
variable the convolution is done. The goal is now to express the
filtered coherence function
\begin{equation}\label{eq:defFilteredG}
G_{\fil}(t_1, \ldots ,t_n) = 
\avg{S_{\fil,1}(t_1)\,S_{\fil,2}(t_2)\cdots S_{\fil,n}(t_n)},
\end{equation}
in terms of the unfiltered coherence function
\begin{equation}
G(t_1, \ldots ,t_n) = \avg{S_1(t_1)\, S_2(t_2)\cdots S_n(t_n)}.
\end{equation}
This can be done straightforwardly by substituting
\eqrf{eq:defFilteredSignal} into \eqrf{eq:defFilteredG}.
\begin{equation}
G_{\fil}(t_1, \ldots ,t_n) = \left<{\prod_{i=1}^{n} f_i(t_i) \ast S_i(t_i)}\right>
\end{equation}
Realizing that all convolutions are related to different time
variables one can rearrange this expression as
\begin{multline}
G_{\fil}(t_1, \ldots ,t_n) = \\
f_1(t_1) \ast  f_2(t_2) \ast  \cdots 
f_n(t_n) \ast G(t_1,\ldots,t_n).
\end{multline}
The integral form clarifies this expression
\begin{multline}
G_{\fil}(t_1, \ldots ,t_n) = \\
\int_{-\infty}^{+\infty}\!\!d\tau_1\cdots\!\!\int_{-\infty}^{+\infty}\!\!d\tau_n\,f_1(t_1 - \tau_1) \cdots f_n(t_n - \tau_n) 
G(\tau_1,\ldots,\tau_n).
\end{multline}
This expression can be seen as a generalized convolution with respect
to more than one time variable. Introducing the global
filter function
\begin{equation}
F(t_1, \ldots ,t_n) = f_1(t_1)\, f_2(t_2)\cdots f_n(t_n),
\end{equation}
one can write 
\begin{equation}
G_{\fil}(t_1, \ldots ,t_n) = F(t_1, \ldots ,t_n) \ast G(t_1, \ldots ,t_n).
\end{equation}

In frequency domain, the same fact can be expressed by using the
multi-dimensional Fourier transform instead, so that one may simply write 
\begin{equation}
G_{\fil}(\omega_1,\ldots,\omega_n) = G(\omega_1,\ldots,\omega_n) 
F(\omega_1,\ldots,\omega_n).
\end{equation}

\subsection{Two-point correlation functions}

Using the spectral representation of some first-order coherence
function $G(t_1,t_2)$ and the global filter function $F(t_1,t_2)$, one
can write $G_{\fil}(\tau)$ as
\begin{align}
G_{\fil}(\tau)
& = \int_{\mathbb R^3} dt\,d\omega_1\,d\omega_2\,\e^{i(\omega_1+\omega_2) t + i\omega_2\tau}
F(\omega_1,\omega_2) G(\omega_1,\omega_2),\\
& = {1\over 2\pi} \int_{-\infty}^{+\infty} d\omega\,\e^{i\omega \tau} 
F(-\omega,\omega) G(-\omega,\omega).
\end{align}
Considering the time representation of this expression, it is clear
that the correlation function will be distorted by a convolution with
the effective two-point correlation function $f_\eff(\tau) = {\mathcal
  F}\{F(-\omega,\omega)/2\pi\}$. Due to the linearity of the filters,
one finds that Dirac $\delta$ in the noise correlations are
replaced by $f_\eff$, so that, for example
\begin{gather}
\int_{\mathcal I}dt\,\avg{\op{h}_c^\dag(t)\op{h}_c(t+\tau)}_\fil 
= \N_c f_\eff(\tau),
\end{gather}
where $h^{(\dag)}_c(t)$ have been introduced in \eqrf{eq:camp} and 
$\avg{\cdot}_\fil$ indicates that the average is taken over
filtered outputs. This illustrates why the values for the different
second order coherence functions remain finite. Moreover, the other
time-integrated two-point correlations are replaced by the convolution
of the two-point correlation function with the $f_\eff$.

\begin{figure}
\includegraphics[width=0.45\textwidth]{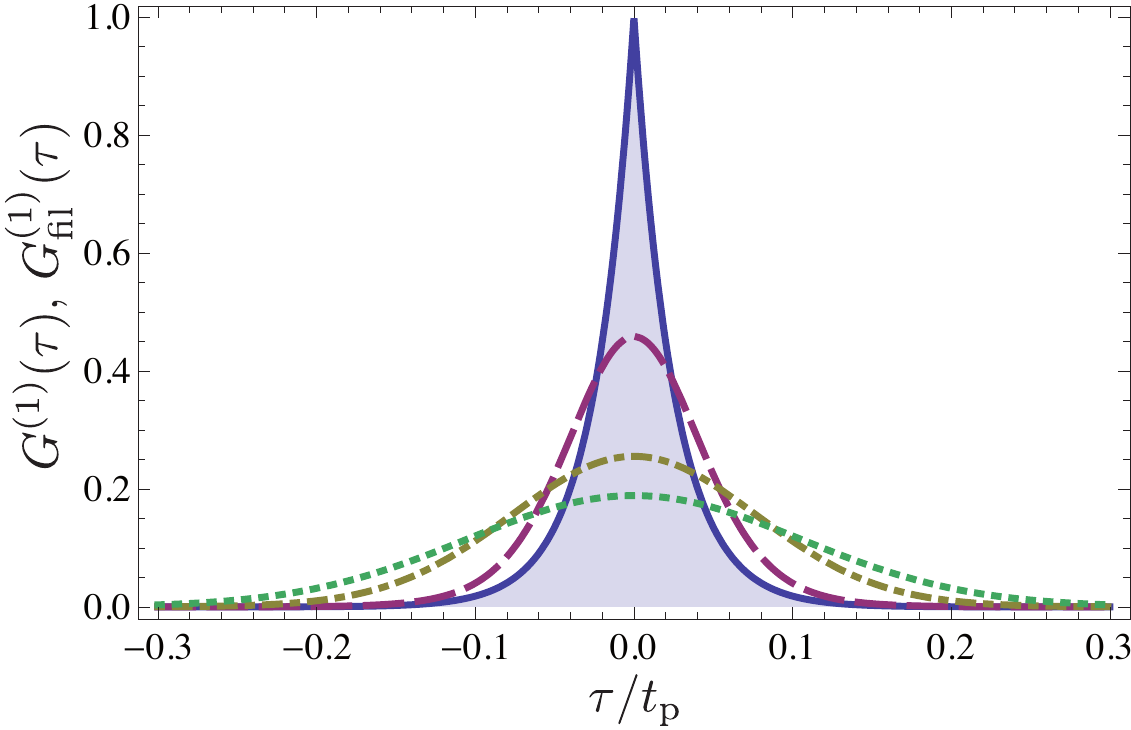}
\caption{(Color online) Distortion of the $\Pi$ pulses in $G^{(1)}(\tau)$ due to
  Gaussian filters of different bandwidths. The solid blue line is
  the unfiltered function, and the others are filtered 
  bandwidth decreasing progressively: 31/$t_\mathrm{p}$ for the
  dashed purple line, 14/$t_\mathrm{p}$ for the dot-dashed yellow line, and
  10/$t_\mathrm{p}$ for the dotted green line.
  \label{fig:g1-pulse-dist}}
\end{figure}

\begin{figure}
\includegraphics[width=0.45\textwidth]{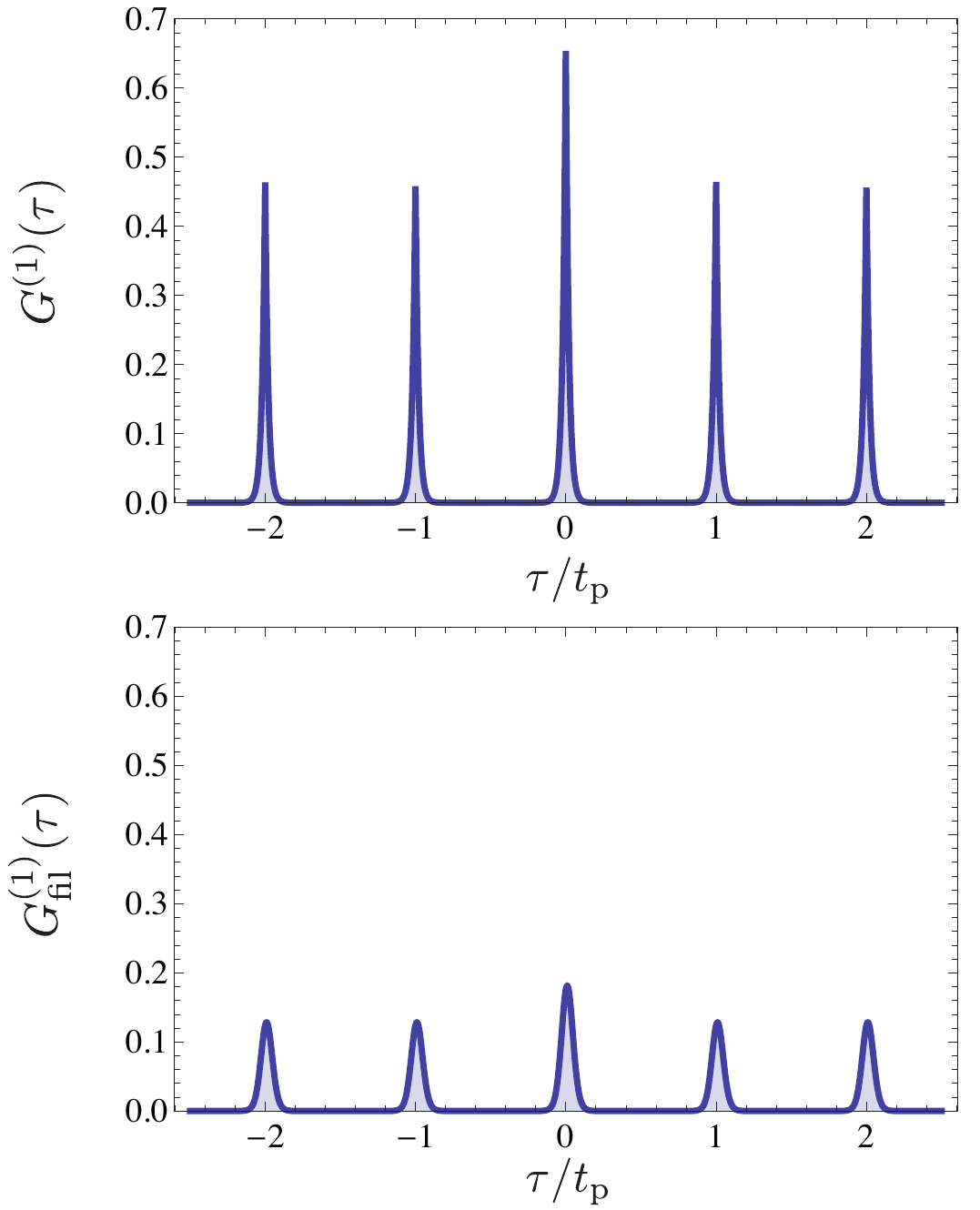}
\caption{(Color online) The pulse train for unfiltered $G^{(1)}(\tau)$ (top) and
  $G^{(1)}_{\fil}(\tau)$ after Gaussian filtering (bottom) under the
  periodic preparation of
  $\sqrt{\sfrac{1}{3}}\ket{0}+\sqrt{\sfrac{2}{3}}\ket{1}$. The
  Gaussian filter used here has a bandwidth of
  $31/\tp$.\label{fig:g1-pulse-train}}
\end{figure}

Note that since a linear time independent filter is
used, the relative heights of the peaks remain unchanged -- only their
shape gets distorted and scaled. This is illustrated in
Fig.~\ref{fig:g1-pulse-dist} and Fig.~\ref{fig:g1-pulse-train}.

\subsection{Four-point correlation functions}

Considering the second-order coherence function with filtered signals, one obtains
\begin{multline}
\GTwofil(\tau) =
\int\limits_{{\mathbb R}^5} dt\,d\omega_1\cdots d\omega_4
\e^{i(\omega_1+\omega_2+\omega_3+\omega_4)t}\e^{i(\omega_2+\omega_4)\tau}\\ 
F(\omega_1,\omega_2,\omega_3,\omega_4) 
\GTwo{}(\omega_1,\omega_2,\omega_3,\omega_4).
\end{multline}
The different types of correlations discussed simply
determine the labeling of the variables. Applying a change of
variables and integrating over time, one obtains
\begin{multline}
\GTwofil(\tau) = 
{1\over8\pi}\int_{-\infty}^{+\infty}\,d\Omega_2\,d\Omega_3\,d\Omega_4\,\e^{-i\Omega_2\tau} \\
F\left({-\Omega_2+\Omega_4\over2},{\Omega_2+\Omega_3\over2},\right.
\left.{\Omega_2-\Omega_3\over2},{-\Omega_2-\Omega_4\over2}\right)\\ 
\GTwo\left({-\Omega_2+\Omega_4\over2},{\Omega_2+\Omega_3\over2},\right.
\left.{\Omega_2-\Omega_3\over2},{-\Omega_2-\Omega_4\over2}\right).
\end{multline}
Note that this leads different behavior, in the sense
that the correlation function is not simply convolved with an
effective impulse response.

\section{Commutation relations of two-sided cavity
  outputs\label{app:commutation}}

Taking both cavity mirrors to be leaky, one finds an additional 
boundary condition in the input-output description of the
cavity ~\cite{collett:1984a,gardiner:1985a,gardiner:2004a,walls:2008a}
\begin{equation}
\op{c}_\out = \sqrt{\kappa_c}\op{a} - \op{c}_\inp,\label{eq:other-bc}
\end{equation}
where the $\op{c}_{\inp,\out}$ mode are now the modes coupling to the
second leaky mirror. The equation of motion \eqrf{eq:aINOUT} for $\op{a}$ in the
rotating frame then becomes
\begin{equation}
\dot{\op{a}} = - {\kappa_b+\kappa_c\over 2}\op{a} +
\sqrt{\kappa_b}\op{b}_\inp + \sqrt{\kappa_c}\op{c}_\inp.
\end{equation}
From Eqs.~\eqref{eq:one-sided-bc} and \eqref{eq:other-bc}, one finds
\begin{multline}
[\op{b}_\out(t),\op{c}^\dag_\out(t')] =
\sqrt{\kappa_b\kappa_c}[\op{a}(t),\op{a}^\dag(t')]\\
-\sqrt{\kappa_b}[\op{a}(t),\op{c}^\dag_\inp(t')]
-\sqrt{\kappa_c}[\op{b}_\inp(t),\op{a}^\dag(t')],
\end{multline}
where the input field operators were taken to commute. Integrating the
solution for the equations of motion of the modes $\op{b}(\omega,t)$
of the left transmission line and the modes $\op{c}(\omega,t)$ of the
right transmission line, and using the definition of the input
fields one obtains
\begin{equation}
\begin{split}
\op{b}_\inp(t) = - {\sqrt{\kappa_b}\over2}\op{a}(t) + 
{1\over\sqrt{2\pi}}\int_{-\infty}^{+\infty}d\omega~\op{b}(\omega,t),\\
\op{c}_\inp(t) = - {\sqrt{\kappa_c}\over2}\op{a}(t) +
{1\over\sqrt{2\pi}}\int_{-\infty}^{+\infty}d\omega~\op{c}(\omega,t).
\end{split}
\end{equation}
From causality and the
boundary conditions above, one finds~\cite{gardiner:2004a}
\begin{align}
  [\op{a}(t),\op{b}_\inp(t')]&=0, &
  [\op{a}(t),\op{c}_\inp(t')]&=0 && \text{ for $t'>t$} \\
  [\op{a}(t),\op{b}_\out(t')]&=0, &
  [\op{a}(t),\op{c}_\out(t')]&=0 && \text{ for $t'<t$}
\end{align}
Combining these commutation relations with the input field
definitions, one finally finds
\begin{equation}
\begin{split}
[\op{a}(t),\op{c}_\inp^\dag(t')] 
&= \sqrt{\kappa_c} u(t-t')[\op{a}(t),\op{a}^\dag(t')],\\
[\op{b}_\inp(t),\op{a}^\dag(t')] 
&= \sqrt{\kappa_b} u(t'-t)[\op{a}(t),\op{a}^\dag(t')],\\
\end{split}
\end{equation}
where
\begin{equation}
u(t) = 
\left\{ 
\begin{array}{lc}
1 & t>0, \\
{1\over2} & t=0,\\
0 & t < 0,
\end{array}
\right.
\end{equation}
and therefore 
\begin{equation}
[\op{b}_\out(t),\op{c}_\out^\dagger(t')]=0,
\end{equation}
as claimed.

\bibliography{ref.bib}

\end{document}